\documentclass[english,twocolumn,superscriptaddress,showpacs,aps,prb,floatfix,reprint,nobibnotes,longbibliography]{revtex4-2}

\usepackage[utf8]{inputenc}
\usepackage{graphicx,subfigure}     
\usepackage{bm}             
\usepackage{verbatim}
\usepackage{epsf}
\usepackage{amssymb,stmaryrd}   
\usepackage{amsmath,amsfonts}
\usepackage{dcolumn}        
\usepackage{siunitx}
\usepackage{xcolor}
\usepackage{float}
\usepackage{amsthm}%
\usepackage[title]{appendix}%
\usepackage{booktabs}%
\usepackage{algorithm}%
\usepackage{algorithmicx}%
\usepackage{algpseudocode}%
\usepackage{listings}%
\usepackage{epstopdf}
\usepackage{natbib}
\usepackage{mathrsfs}
\usepackage{xspace}
\usepackage{placeins}
\usepackage{babel}
\graphicspath{ {images/} }  
\usepackage{wrapfig}
\usepackage{tikz}  
\usetikzlibrary{arrows,shapes,trees,positioning}  
\usepackage{comment}
\usepackage{textcomp}

\usepackage[colorlinks=true,citecolor=blue]{hyperref}

\usepackage{soul}
\setstcolor{red}
\usepackage{ulem}
\hypersetup{
	colorlinks=true,
	linkcolor=blue,
	citecolor=blue,
	urlcolor=blue,
}
\DeclareUnicodeCharacter{0096}{_}

\DeclareUnicodeCharacter{2212}{\textminus}

\definecolor{viol}{rgb}{0.7, 0.4, 1}

\definecolor{mygray}{cmyk}{0, 0, 0, 0.3}

\newcommand{\MMS}{MgMn$_6$Sn$_6$\xspace}

\begin{document}

\title{Ferromagnetic Resonance Spectroscopy on the Kagome Magnet \MMS}

\author{Riju Pal}
\email{rijupal07@gmail.com}
\affiliation{Leibniz Institute for Solid State and Materials Research, Helmholtzstr. 20, D-01069, Dresden, Germany}

\author{Kakan Deb}
\affiliation{Department of Condensed Matter and Materials Physics, S. N. Bose National Centre for Basic Sciences, Block JD, Sector III, Salt Lake, Kolkata, 700106, India}

\author{Nitesh Kumar}
\affiliation{Department of Condensed Matter and Materials Physics, S. N. Bose National Centre for Basic Sciences, Block JD, Sector III, Salt Lake, Kolkata, 700106, India}

\author{Bernd Büchner}
\affiliation{Leibniz Institute for Solid State and Materials Research, Helmholtzstr. 20, D-01069, Dresden, Germany}
\affiliation{Institute for Solid State and Materials Physics and Würzburg-Dresden Cluster of Excellence ctd.qmat, TU Dresden, D-01062, Dresden, Germany}

\author{Alexey Alfonsov}
\affiliation{Leibniz Institute for Solid State and Materials Research, Helmholtzstr. 20, D-01069, Dresden, Germany}

\author{Vladislav Kataev}
\email{v.kataev@ifw-dresden.de}
\affiliation{Leibniz Institute for Solid State and Materials Research, Helmholtzstr. 20, D-01069, Dresden, Germany}

\date{\today}

\begin{abstract}

\MMS is the itinerant ferromagnet on the kagome lattice with high ordering temperature featuring complex electronic properties due to the nontrivial topological electronic band structure, where the spin-orbit coupling (SOC) plays a crucial role. Here, we report a detailed ferromagnetic resonance (FMR) spectroscopic study of \MMS aimed to elucidate and quantify the intrinsic magnetocrystalline anisotropy that is responsible for the alignment of the Mn magnetic moments in the kagome plane. By analyzing the frequency, magnetic field, and temperature dependences of the FMR modes, we have quantified the magnetocrystalline anisotropy energy density that reaches the value of approximately $ 3.5\cdot 10^6$\,erg/cm$^3$ at $T = 3$\,K and reduces to about $1\cdot 10^6$\,erg/cm$^3$ at $T = 300$\,K. The revealed significantly strong magnetic anisotropy suggests a sizable contribution of the orbital magnetic moment to the spin magnetic moment of Mn, supporting the scenario of the essential role of SOC for the nontrivial electronic properties of \MMS.\\

\end{abstract}

\maketitle

\section{Introduction}\label{sec:Introduction}
\begin{figure*}
	\centering
	\includegraphics[width=0.75\linewidth]{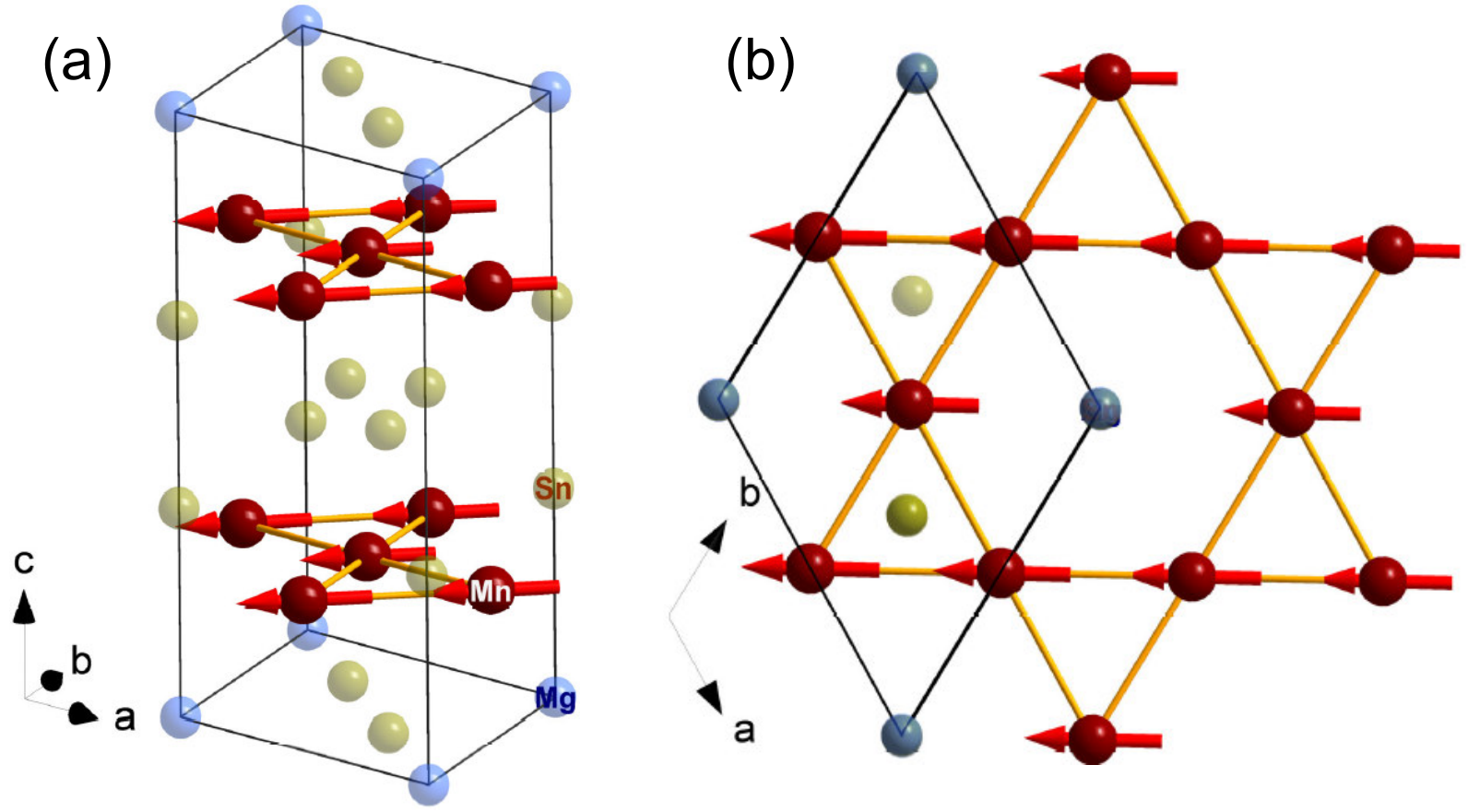}
	\caption{Crystal structure of \MMS. The atoms are colored as Mg = light blue, Mn = red, Sn = dark yellow. Red arrows depict the direction of the Mn magnetic moments in the ferromagnetically ordered state. (a) The unit cell can be viewed as composed of two slabs of atoms Mn--(Mg-Sn)--Mn and Mn--Sn--Sn--Sn--Mn stacking along the $c$-axis. (b) Mn atoms in each slab form the kagome lattice with the magnetic moments ferromagnetically ordered in the $ab$-plane.} 
	\label{fig:structure}
\end{figure*}

Kagome magnets constitute a broad class of materials featuring the so-called kagome lattice composed of hexagons connected with each other by equilateral triangles (Fig.~\ref{fig:structure}). The triangular nature of this lattice implies strong geometrical frustration, which in the case of antiferromagnetic interactions may give rise to a strongly correlated spin liquid state. Its electronic structure features non-trivial topology with Dirac, Weyl, and nodal line points, van Hove singularities, and flat bands potentially giving rise to unconventional quantum phenomena such as unconventional superconductivity, charge density waves, and the anomalous Hall effect (for recent reviews see, e.g., Refs.~\cite{Yin2022,Wang2023,Li2025}). One of the interesting subclasses of the kagome magnets that belong to the so-called `166' family \cite{Xu2023} are compounds with the general chemical composition $R$Mn$_6$Sn$_6$ with $R$ =  Mg, Sc, Y, Zr, Hf, Gd, Tm, Lu. Depending on the type of the $R$ element, they demonstrate different magnetic ground states, ranging from collinear ferro-, antiferro-, and ferrimagnetic orders to helimagnetic order \cite{Chafik1991,Venturini1991,Venturini1993,Venturini1996,Mazet1998,Mazet1999a,Mazet1999b}. One particular material in this series, the itinerant ferromagnet \MMS,  has recently received much attention due to the computational predictions of the non-trivial topology of the electronic structure hosting Dirac fermions, van Hove singularities, and flat bands near the Fermi energy, and anomalous Hall conductivity \cite{Sau2024}. Indeed, a significant anomalous Hall effect was very recently observed in the transport measurements on \MMS single crystals \cite{Ma2025}. 

Regarding magnetism, the static magnetic properties of \MMS were studied in quite detail \cite{Mazet1998,Mazet1999a,Song2024a,Song2024b}. This compound adopts the HfFe$_6$Ge$_6$ type of structure P6/$mmm$ (No. 191) featuring in the unit cell two distinct slabs of atoms Mn--(Mg-Sn)--Mn and Mn--Sn--Sn--Sn--Mn that stack along the $c$-axis [Fig.~\ref{fig:structure}(a)] \cite{Mazet1998}. Magnetic Mn atoms form the kagome lattice in each slab, where the Mn spins are coupled ferromagnetically. According to the neutron diffraction data, below $T_{\rm C}\approx 300$\,K 
%
%
the spins order ferromagnetically in the basal plane [Fig.~\ref{fig:structure}(b)] such that the magnetic structure consists of ferromagnetic Mn(001) planes ferromagnetically coupled along the $c$-axis [Fig.~\ref{fig:structure}(a)] \cite{Mazet1999a}. 
%

%

However, one important aspect of magnetism of \MMS, the origin and quantification of magnetic anisotropy, which dictates the in-plane alignment of Mn magnetic moments, was not addressed so far. To fill this gap, we conducted in this work detailed ferromagnetic resonance (FMR) measurements on \MMS in the broad ranges of excitation frequencies, magnetic fields, and temperatures. Since the properties of the FMR signal sensitively depend on the magnetic anisotropy of a ferromagnet, we could make reliable estimates of the intrinsic magnetocrystalline anisotropy energy density of \MMS from the analysis of the frequency and temperature dependences of the FMR modes for different orientations of the applied magnetic field with respect to crystallographic directions. We have found that the uniaxial magnetocrystalline anisotropy $K_{\rm uniax}$ is positive, i.e., of the easy-plane type, and is significant, amounting to $K_{\rm uniax} \approx 3.5\cdot 10^6$\,erg/cm$^3$ at the lowest temperature. This finding suggests an important role of the spin-orbit coupling (SOC) in defining the ground state spin structure of \MMS in line with the predictions of the crucial role of SOC for the topological electronic band structure and anomalous Hall effect \cite{Sau2024,Li2025}.  

\section{Experimental details}
\label{sec:details}

High-quality single crystals of \MMS were synthesized using the self-flux method \cite{Song2024a}. Mg turnings (99.9\%, Alfa Aesar), Mn pieces (99.9\%, Alfa Aesar), and Sn pieces (99.98\%, Alfa Aesar) were mixed in an atomic ratio of 4:1:8 and loaded into an alumina crucible. The crucible was then sealed in a quartz ampule under high vacuum ($< 1 \times 10^{-5}$ mbar). The sealed ampule was heated in a muffle furnace to 1073~K, held for 10~hours to ensure homogenization, and then slowly cooled to 693~K at a rate of 4~K~h$^{-1}$. At this temperature, hexagonal plate-like single crystals were obtained after removing the excess flux via centrifugation. Their analytical characterization (composition and crystal structure) as well as static magnetic properties are described in the Appendix.

FMR measurements were carried out with a home-made multi-frequency electron spin resonance spectrometer. For the generation and detection of microwaves in the frequency range 75--330\,GHz, a vector network analyzer (PNA-X) from Keysight Technologies with the extensions from Virginia Diodes, Inc. was employed. The incident and transmitted microwaves were guided to the sample and back to the analyzer using a probehead with oversized waveguides. The probehead was placed in the superconducting magnet system from Oxford Instruments that enabled magnetic field sweeps in the range 0--16\,T, as well as variation and stabilization of the sample's temperature in the range 3--300\,K. The unavoidable mixing of the absorption and dispersion components of the detected signal due to the complex impedance of the broadband probehead was accounted for on the software level, yielding correct values of the resonance field and linewidth of the measured signals (for details see \cite{Pal2024, Abraham2023}). 

\section{Computational background}
\label{sec:SpinW}

\begin{figure*}[ht]
	\centering
	\includegraphics[width=1.0\linewidth]{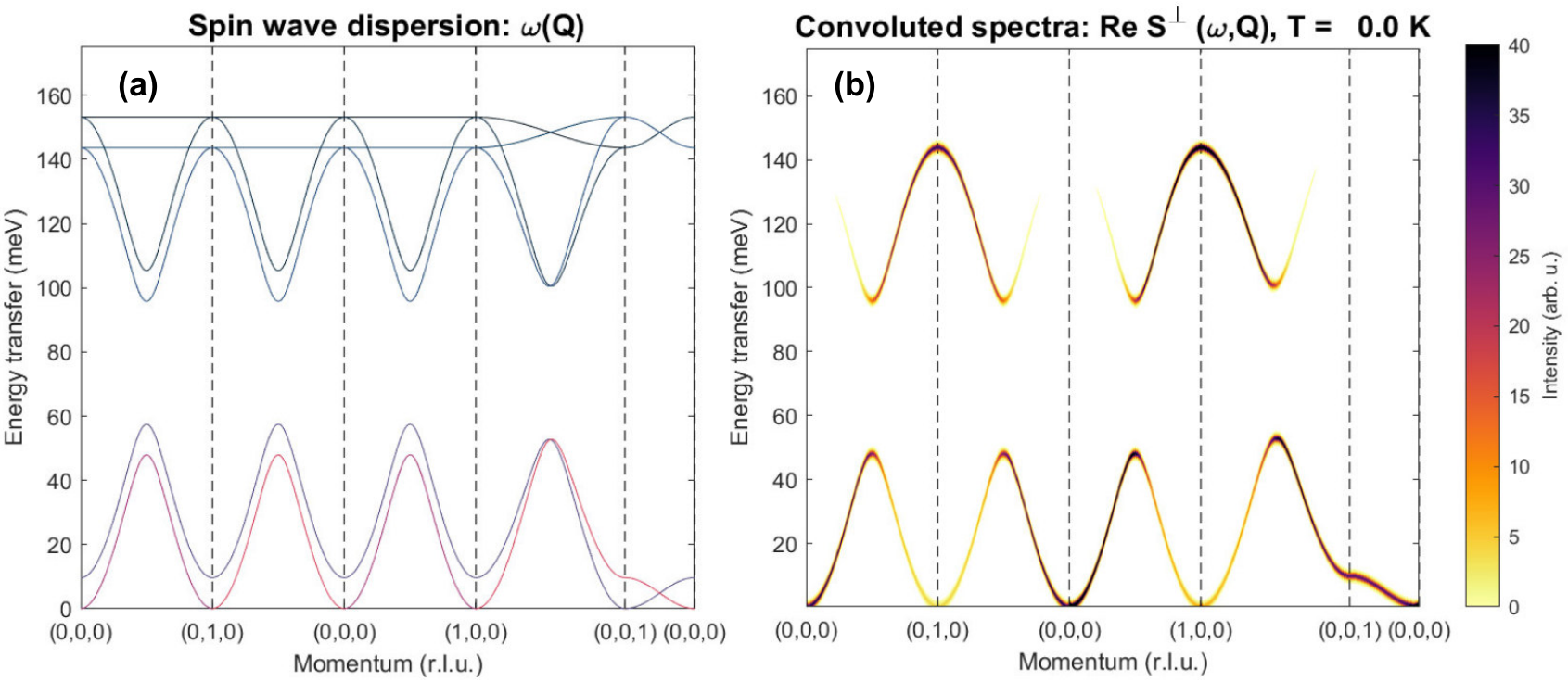}
	\caption{Spin wave excitation spectrum of \MMS modeled with the SpinW software. (a) Energy dispersion of the 6 spin wave modes in the magnetic Brillouin zone. (b)  Inelastic neutron scattering intensity (cross-section) Re\,S$^\perp(\hbar\omega,\mathbf{q}$) as a function of the momentum transfer $\mathbf{q}$.}
	\label{fig:SW}
\end{figure*}

The excitation spectrum of ferromagnetic spin waves in \MMS is expected to have six modes corresponding to the six magnetic Mn atoms in the unit cell. The spectrum can be modeled with the SpinW software used for the analysis of the inelastic neutron scattering experiments \cite{Toth_2015,SpinW} (Fig.~\ref{fig:SW}). In this approach, the following general magnetic Hamiltonian of interacting localized magnetic moments on a periodic lattice is solved using the linear spin wave theory \cite{Toth_2015}:
\begin{eqnarray}
	\mathcal{H} = \sum_{i,j}\mathbf{S}_iJ_{i,j}\mathbf{S}_j + \sum_{i}\mathbf{S}_iA\mathbf{S}_i + \mu_{\rm B}\sum_{i}\mathbf{H}g\mathbf{S}_i 
\label{eq:hamilton}
\end{eqnarray}
Here, $\mathbf{S_{i,j}}$ are vectors of the spin operators, $\mathbf{H}$ is the magnetic field vector, $J$ and $A$ are tensors of the exchange interaction and of the single ion anisotropy, respectively, and $g$ is the $g$-factor tensor.

In the case of \MMS, we assume for simplicity the isotropy of the exchange interaction and of the $g$-factor, taking the latter equal to the spin-only value of 2. The result of the modeling is shown in Fig.~\ref{fig:SW}. The intra-planar ferromagnetic exchange interaction constant $J_{\rm intra}$ was taken as  the mean-field estimate from the Curie-Weiss temperature $\Theta_{\rm CW} = 370$\,K \cite{Mazet1999a}, yielding $J_{\rm intra} = - 3|\Theta_{\rm CW}|/zS(S+1) \approx -111\,{\rm K} = - 9.57$\,meV. Here, $z = 4$ is the number of nearest neighbors in the kagome lattice. The effective spin $S$ of the Mn atom was estimated from the ordered moment $\mu_{\rm Mn} = gS\mu_{\rm B} = 2.32\mu_{\rm B}$ \cite{Mazet1998,Mazet1999a} with $g = 2$. The likewise ferromagnetic inter-plane constant was set as $J_{\rm inter} = 0.1J_{\rm intra}$.

\begin{figure}
	\centering
	\includegraphics[width=\linewidth]{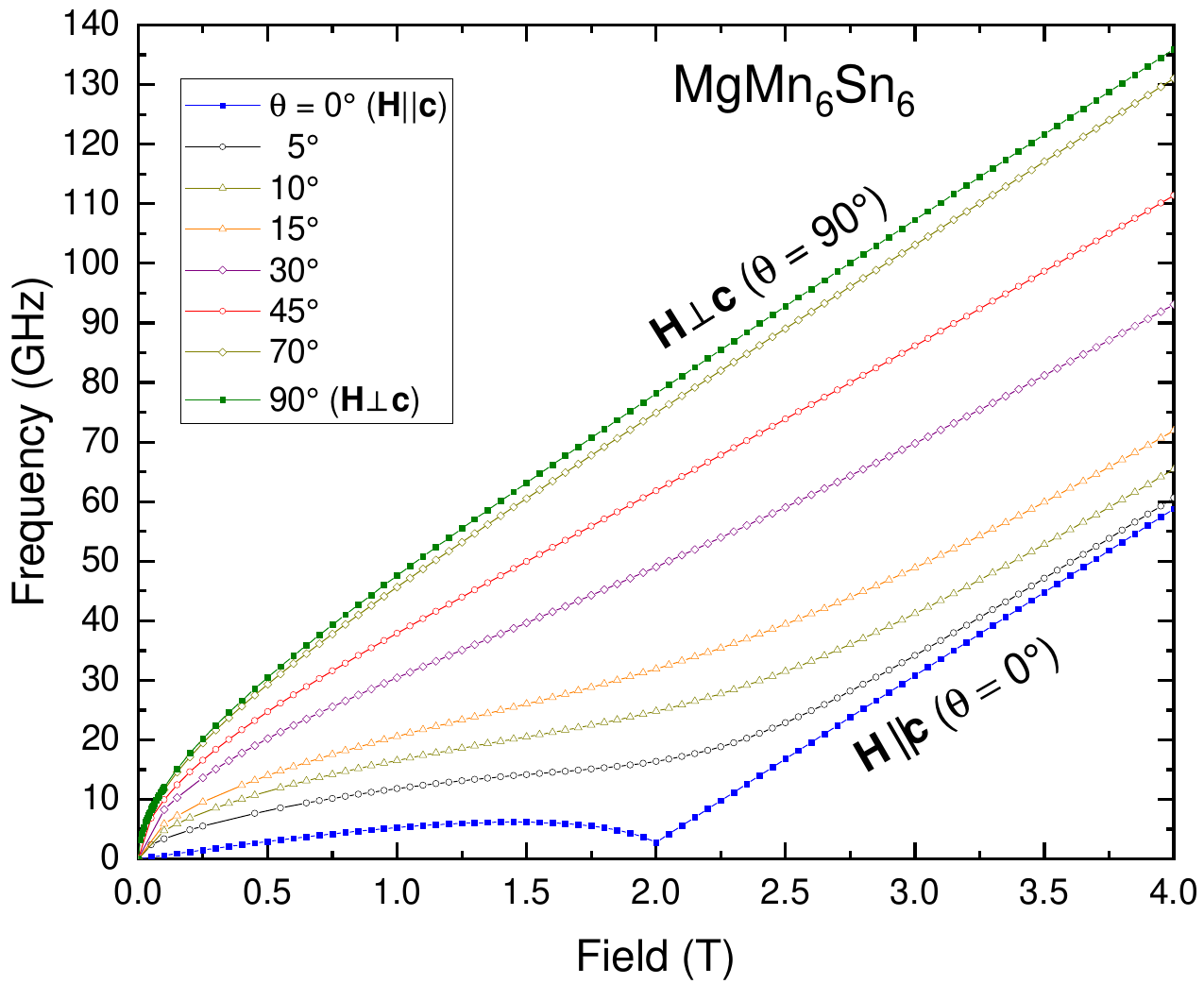}
	\caption{Magnetic field dependence of the energy of the uniform (${\bf q} = 0$) spin wave mode of \MMS for different orientations of the applied magnetic field calculated with the SpinW software.}
	\label{fig:SW_field}
\end{figure}

While Fig.~\ref{fig:SW}(a) shows the dispersion of the spin wave modes in the magnetic Brillouin zone, Fig.~\ref{fig:SW}(b) depicts the neutron magnetic scattering intensity of these modes, which is proportional to the imaginary part of the dynamical magnetic susceptibility probed in the  ESR absorption measurements. Given that ESR is limited to the wave-vector ${\bf q} = 0$ excitations, it is evident from Fig.~\ref{fig:SW}(b) that only one gapless mode at the (0,0,0) zone center should be observed by FMR in \MMS.

The energy dispersion of this mode in the applied magnetic field is shown in Fig.~\ref{fig:SW_field} for different orientations of the field vector with respect to the crystallographic axes. In this calculation the single ion anisotropy constant of the Mn ions was chosen to be $A^{\rm zz} = 0.1$\,meV to account for the easy-plane type of magnetic anisotropy of \MMS and to reproduce the critical field $H_{\rm a} = 2$\,T at which the experimentally observed FMR mode for $\mathbf{H}\parallel \mathbf{c}$ softens towards zero frequency at the lowest temperature (see Section~\ref{sec:results}).

\FloatBarrier

\section{Experimental results}\label{sec:results}

\begin{figure*}[ht]
	\centering
     \includegraphics[width=1.0\linewidth]{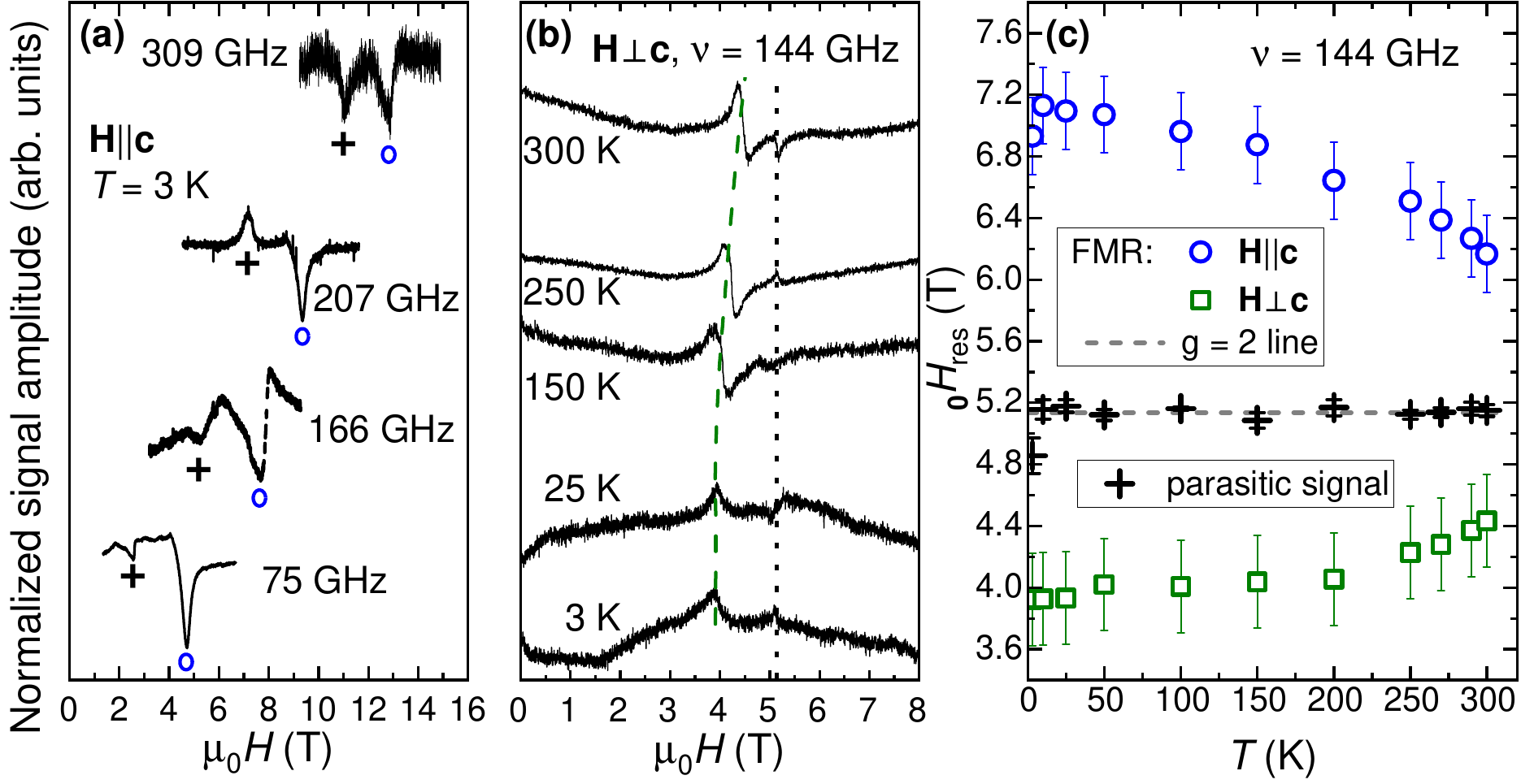}
	\caption{(a) Spectra at different frequencies for $\mathbf{H}\parallel\mathbf{c}$ at $T = 3$\,K. The FMR signal is marked by circles, and the parasitic line is marked by crosses. (b) Spectra at different temperatures for $\mathbf{H}\perp\mathbf{c}$ at $\nu = 144$\,GHz. The position of the FMR signal is traced by the dashed line, and that of the parasitic signal is marked by the vertical dotted line at the resonance field corresponding to $g = 2$ at this frequency. 
	(c) Temperature dependence of the resonance field of the FMR signal for $\mathbf{H}\parallel\mathbf{c}$ (circles) and $\mathbf{H}\perp\mathbf{c}$ (squares), and of the parasitic signal (crosses) at $\nu = 144$\,GHz. The dashed line denotes the paramagnetic resonance field with $g = 2$ at this frequency.}
	\label{fig:HF_ESR1}
\end{figure*}

Exemplary spectra at selected frequencies measured at $T = 3$\,K for $\mathbf{H}\parallel\mathbf{c}$ field geometry are presented in Fig.~\ref{fig:HF_ESR1}(a). Besides the FMR signal denoted in this plot by circles, there is an additional parasitic line marked by crosses. Its position at a given frequency does not depend on temperature and orientation of the sample. Fig.~\ref{fig:HF_ESR1}(b) shows exemplary spectra at selected temperatures measured at a fixed frequency $\nu = 144$\,GHz for $\mathbf{H}\perp\mathbf{c}$ configuration. The FMR signal (left in the spectrum) shifts from lower to higher fields with increasing the temperature, whereas the parasitic signal always stays at the constant field $H_{\rm res} = h\nu/g\mu_{\rm B}\mu_{0}$ corresponding to the $g$-factor of 2. This field is indicated in Fig.~\ref{fig:HF_ESR1}(c) by the horizontal dashed line. Thus, this signal can be tentatively ascribed to the presence in the sample of spurious magnetic centers of unidentified origin with the $g$-factor very close to 2. In contrast, as can be seen in Fig.~\ref{fig:HF_ESR1}(c), the FMR signals are strongly shifted from the paramagnetic line to higher fields for $\mathbf{H}\parallel\mathbf{c}$ and to lower fields for $\mathbf{H}\perp\mathbf{c}$. These shifts are progressively decreasing with rising temperature towards the ordering temperature $T_{\rm C} \approx 298$\,K.

The $\nu(H_{\rm res})$ dependences of the FMR signal (FMR branches) measured at $T = 3$\,K  for both field orientations are presented in Fig.~\ref{fig:HF_ESR2}. As can be seen there, they are asymmetrically shifted with respect to the isotropic paramagnetic (PM) branch (dashed line). The $\nu(H_{\rm res})$ dependence of the parasitic signal (crosses) is isotropic and it closely follows this line. The FMR branches for both field directions can be successfully numerically modeled with the SpinW software. The results are plotted in Fig.~\ref{fig:HF_ESR2} by solid lines. 

As the temperature increases, the shifts of the FMR signal from the paramagnetic resonance position gradually decrease but still remain finite at 300\,K, i.e., slightly above the ordering temperature $T_{\rm C}$ [Fig.~\ref{fig:HF_ESR1}(c)]. This is also clearly seen in the $\nu(H_{\rm res})$ diagram of the FMR modes measured at $T = 300$\,K where the FMR branches moved closer to the paramagnetic branch, as compared to Fig.~\ref{fig:HF_ESR2}, but still did not merge with it (Fig.~\ref{fig:HF_ESR3}). These results indicate that at least short-range FM order is still present in \MMS at this temperature. 

\FloatBarrier

\section{Discussion}
\label{sec:discussion}

\begin{figure*}
	\centering
	\includegraphics[width=0.65\linewidth]{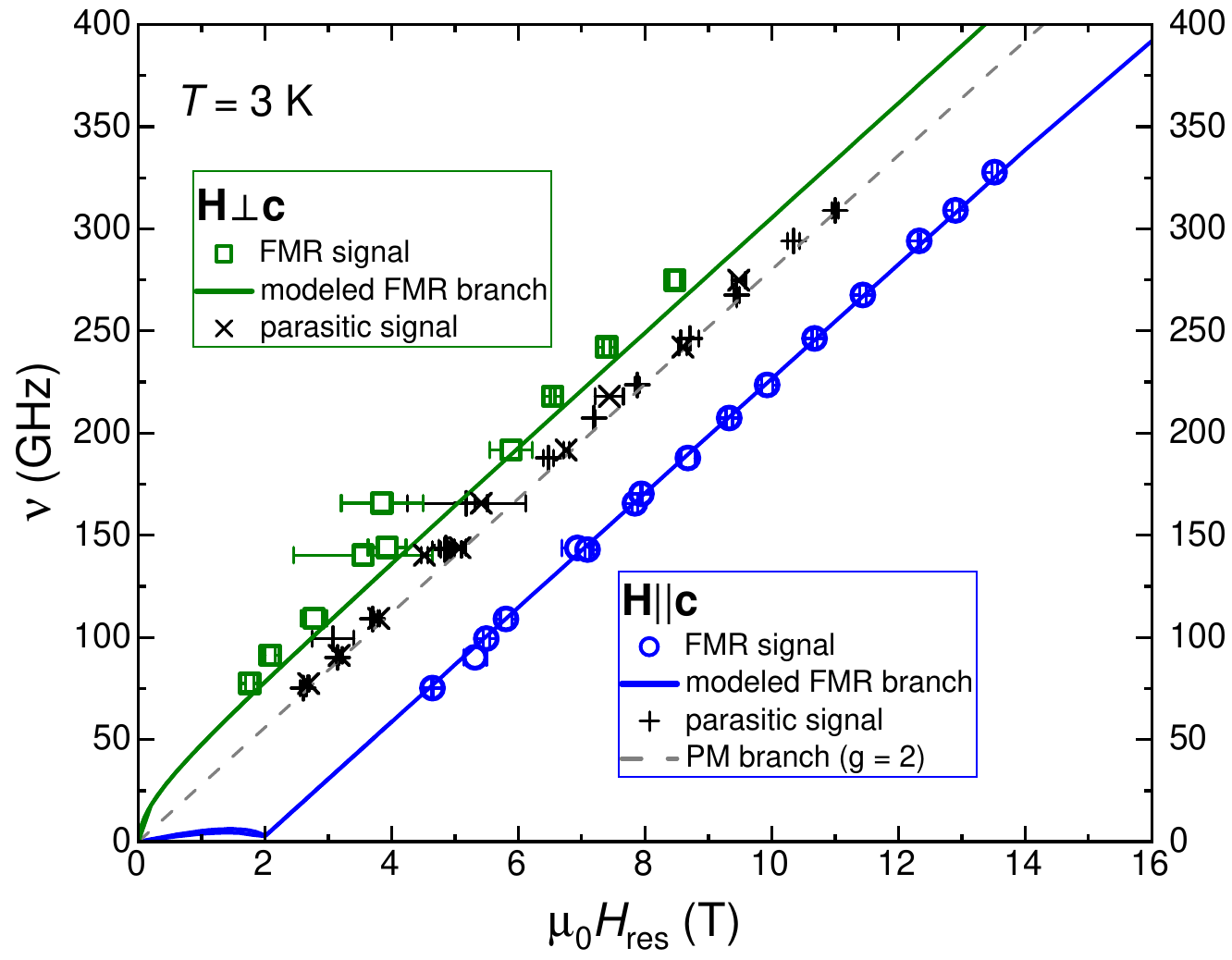}
	\caption{Frequency $\nu$ {\it versus} resonance field $H_{\rm res}$ dependence of the spectral lines at $T = 3$\,K for $\mathbf{H}\parallel\mathbf{c}$ and $\mathbf{H}\perp\mathbf{c}$ field geometries. Circles and squares correspond to FMR signals in the two field orientations, and crosses denote $H_{\rm res}$ of the parasitic signal. Solid curves are results of the numerical modeling of the FMR branches. Dashed line denotes the paramagnetic resonance branch $\nu = (g\mu_{\rm B}\mu_{0}/h)H$ with $g = 2$.} 
	\label{fig:HF_ESR2}
\end{figure*}

\begin{figure*}
	\centering
	\includegraphics[width=0.65\linewidth]{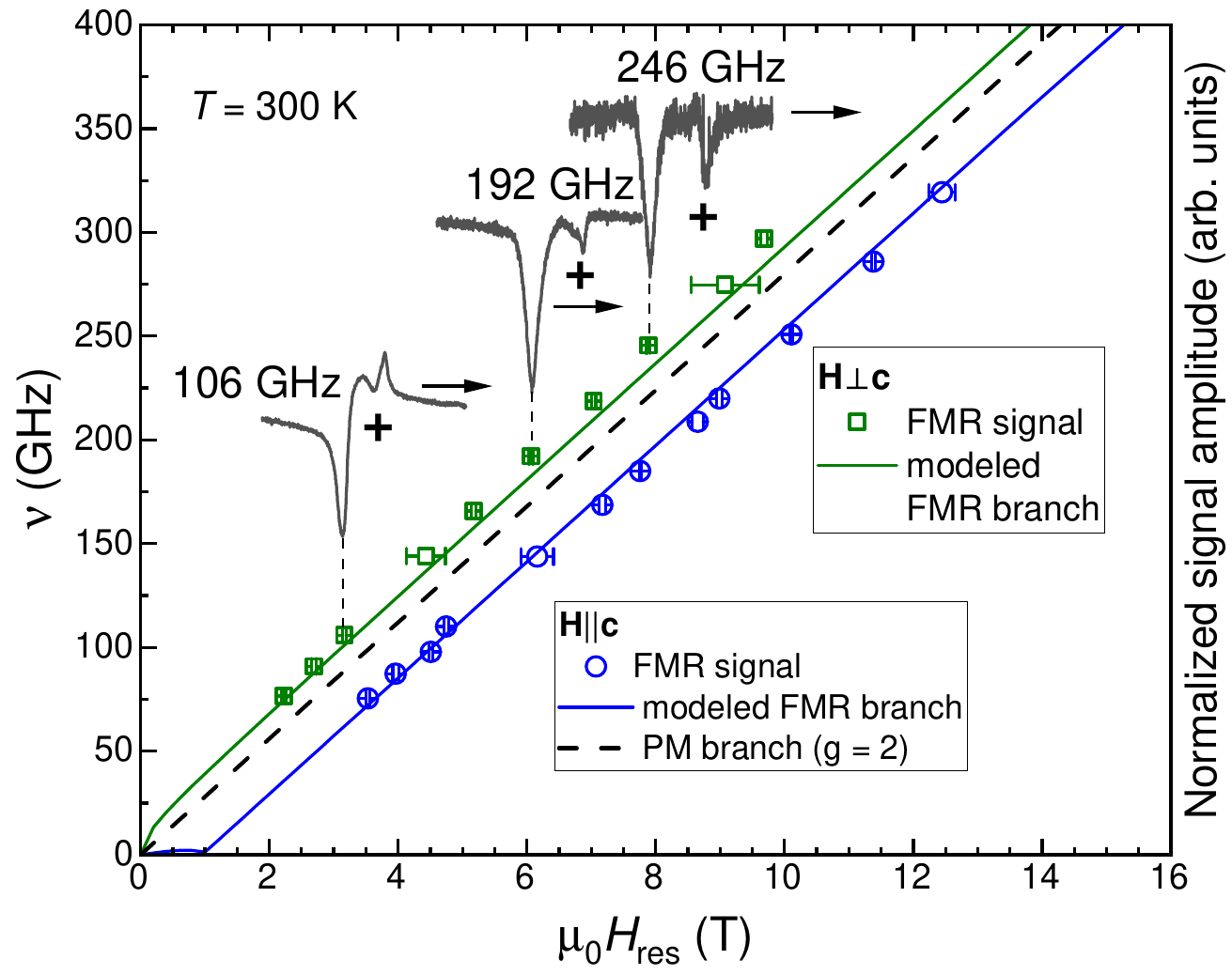}
	\caption{(Left vertical scale) Frequency $\nu$ {\it versus} resonance field $H_{\rm res}$ dependence of the FMR spectral lines at $T = 300$\,K for $\mathbf{H}\parallel\mathbf{c}$ and $\mathbf{H}\perp\mathbf{c}$ field geometries. Circles and squares correspond to FMR signals in the two field orientations. Solid curves are results of the numerical modeling of the FMR branches. Dashed line denotes the paramagnetic resonance branch $\nu = (g\mu_{\rm B}\mu_{0}/h)H$ with $g = 2$. The parasitic signal is not shown here for clarity as it follows closely this paramagnetic branch, similar to Fig.~\ref{fig:HF_ESR2}. (Right vertical scale) Exemplary spectra at different frequencies for $\mathbf{H}\perp\mathbf{c}$ field configuration. The parasitic line in the spectra with $g = 2$ is marked by crosses. }
	\label{fig:HF_ESR3}
\end{figure*}

The frequency and temperature dependence of the anisotropic shifts of the FMR signal presented in Section~\ref{sec:results} evidence the easy-plane character of the ferromagnetic state of \MMS and support the conclusions of the previous static magnetic measurements and magnetic neutron diffraction experiments \cite{Mazet1998,Mazet1999a}. Now, an important question on the origin of the anisotropy, which has not yet been addressed so far, can be answered by the analysis of the FMR data. The $\nu(H_{\rm res})$ FMR branches in Fig.~\ref{fig:HF_ESR2} and Fig.~\ref{fig:HF_ESR3} could be successfully modeled with the SpinW software setting the positive, i.e., easy-plane, single-ion anisotropy ($A^{\rm zz}>0$) in Hamiltonian~(\ref{eq:hamilton}). However, in principle, the shape anisotropy of the plate-like crystals of \MMS could also give rise to the easy-plane anisotropy of the magnetic response and could yield qualitatively the same behavior of the FMR branches. 

To disentangle the two possible sources of magnetic anisotropy, it is instructive to consider the phenomenological model of ferromagnetic resonance based on the Hamiltonian of magnetic free energy density \cite{Turov}:
\begin{equation}
	\mathcal{F} = \mathbf{H \cdot M} + K_\mathrm{\rm uniax} \frac{{M_{\rm z}}^2}{M_{\rm s}^2} + 2\pi M_{\rm z}^2N_{\rm z} \ .
	\label{eq:free_energy}
\end{equation}   

Here, the first term stands for the Zeeman interaction, the second term represents the intrinsic uniaxial magnetocrystalline anisotropy characterized by the energy density $K_{\rm uniax}$, and the third term denotes the shape anisotropy of the plate-like sample with the demagnetization factor $N_{\rm z}$ \cite{Osborn1945,Cronemeyer1991}. $M_{\rm z}$ is the $z$-component of the magnetization vector $\mathbf{M}$ and $M_{\rm s}$ is the saturation magnetization. Based on this Hamiltonian, the linear spin wave theory gives the following analytical solutions for the frequencies of the FMR modes of an easy-plane ferromagnet \cite{Turov}:
\begin{align}
\mathbf{H}\parallel \mathbf{c}: \   & \nu = (g\mu_{\rm B}\mu_{0}/h) (H-H_{\rm a}), & H> H_{\rm a}, \label{eq:freq_c_axis}\\
& \nu = 0, & H \leq H_{\rm a}, \nonumber \\
& & \nonumber \\ 
\mathbf{H}\perp \mathbf{c}: \  &  \nu = (g\mu_{\rm B}\mu_{0}/h) \sqrt{H(H+H_{\rm a})} \ . &  \label{eq:freq_ab_plane}
\end{align}

\begin{figure*}[t]
	\centering
	\includegraphics[width=0.75\linewidth]{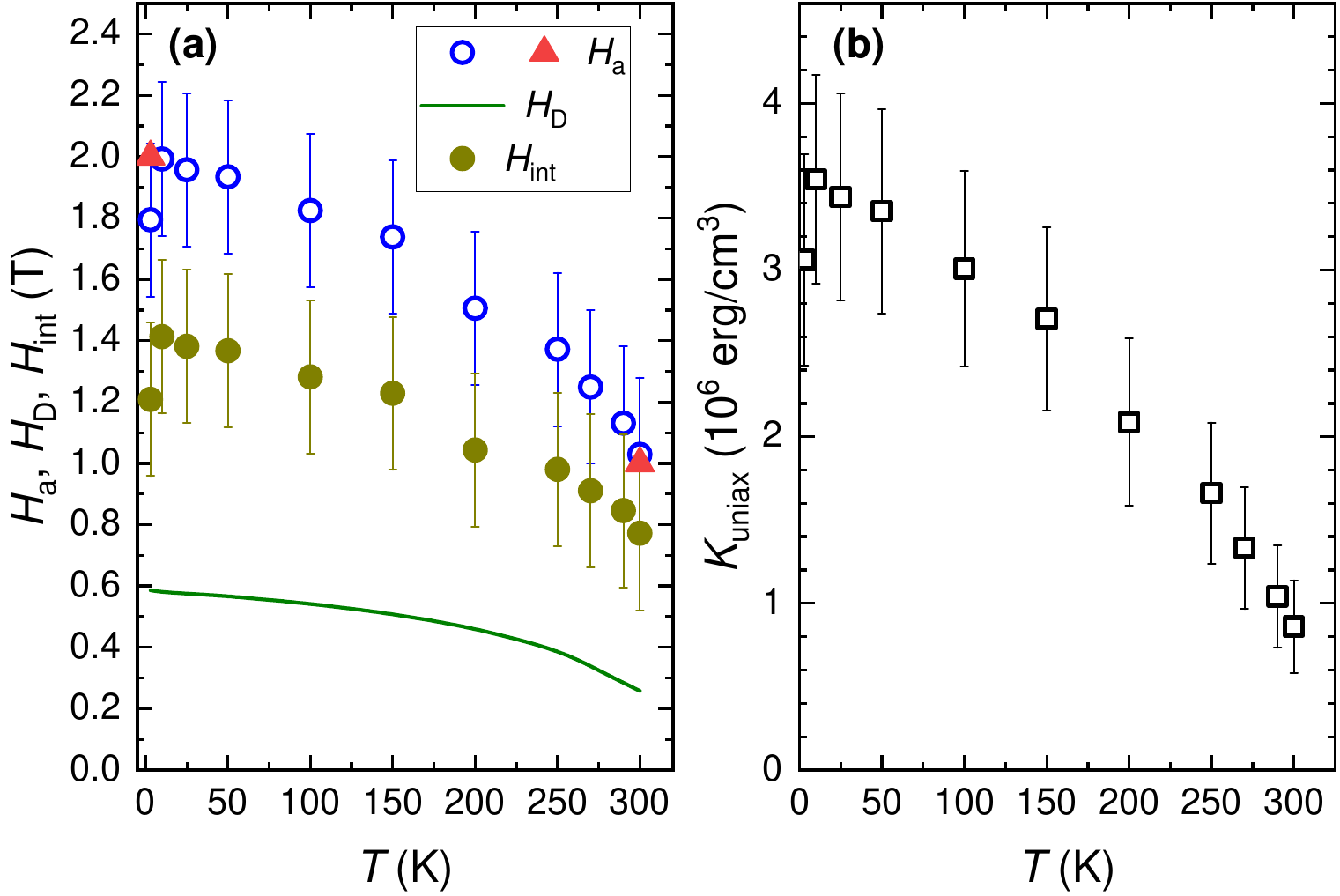}
	\caption{(a) Total anisotropy field $H_{\rm a}$ calculated from the $T$-dependence of the resonance field for $\mathbf{H}\parallel\mathbf{c}$ in Fig.~\ref{fig:HF_ESR1}(c) (open circles) and obtained from the fit of the FMR branches in Fig.~\ref{fig:HF_ESR2} and Fig.~\ref{fig:HF_ESR3} (filled triangles), demagnetization field $H_{\rm D}$ calculated from the $T$-dependence of the saturation magnetization (solid line), and the intrinsic magnetocrystalline anisotropy field $H_{\rm int} = H_{\rm a} - H_{\rm D}$ (filled circles). (b) Temperature dependence of the uniaxial magnetocrystalline energy density $K_{\rm uniax}$.}    
	\label{fig:anisotropy}
\end{figure*}

Here, $H_{\rm a} \geq 0$ is the effective total anisotropy field of a ferromagnet comprising the intrinsic magnetocrystalline field $H_{\rm int}$ and the demagnetization field $H_{\rm D} = 4\pi N_{\rm z}M_{\rm s}$ due to the shape anisotropy, $H_{\rm a} = H_{\rm int} + H_{\rm D} $. Fitting the data in Fig.~\ref{fig:HF_ESR2} and Fig.~\ref{fig:HF_ESR3} with Eqs.~(\ref{eq:freq_c_axis}) and (\ref{eq:freq_ab_plane}) exactly reproduces the SpinW modeling and yields the values of $\mu_{0}H_{\rm a}(T = 3\,{\rm K}) = 2$\,T and $\mu_{0}H_{\rm a}(T = 300\,{\rm K}) = 1$\,T at which the FMR mode for $\mathbf{H}\parallel \mathbf{c}$ softens towards zero frequency. The entire temperature dependence of $H_{\rm a}$ can be estimated from that of the resonance field $H_{\rm res}$ of the FMR signal for $\mathbf{H}\parallel \mathbf{c}$ [Fig.~\ref{fig:HF_ESR1}(c)] by rewriting Eq.~(\ref{eq:freq_c_axis}) as $H_{\rm a}(T) = H_{\rm res}(T) - h\nu/(g\mu_{\rm B}\mu_{0})$ with $\nu = 144$\,GHz and $g = 2$. The result is shown in Fig.~\ref{fig:anisotropy}(a) by open circles. As can be seen there, the $H_{\rm a}(T)$ dependence matches very well with the reference data points $\mu_{0}H_{\rm a}(T = 3\,{\rm K}) = 2$\,T and $\mu_{0}H_{\rm a}(T = 300\,{\rm K}) = 1$\,T obtained from the above-discussed fit of the FMR branches $\nu(H_{\rm res})$ (filled triangles). The demagnetization factor for the particular geometry of the measured plate-like crystal was estimated to be $N_{\rm z} = 0.92$ \cite{Osborn1945,Cronemeyer1991}. This enables a straightforward calculation of the temperature dependence of the demagnetization field $H_{\rm D} = 4\pi N_{\rm z}M_{\rm s}$ shown by the solid line in Fig.~\ref{fig:anisotropy}(a) from the known $T$-dependence of the saturation magnetization (Fig.~\ref{fig:magnetization}). 

It becomes evident from Fig.~\ref{fig:anisotropy}(a)  that the shape anisotropy makes a minor contribution to the total anisotropy field, and the intrinsic magnetocrystalline anisotropy field $H_{\rm int} = H_{\rm a} - H_{\rm D} $ plotted in this Figure by filled circles is significant and dominates in the entire temperature range up to $T_{\rm C}$. The respective intrinsic magnetocrystalline anisotropy energy density $K_{\rm uniax}$ is related to $H_{\rm int}$ as $K_{\rm uniax} = H_{\rm int}M_{\rm s}/2$ \cite{Turov}. Its temperature dependence is shown in Fig.~\ref{fig:anisotropy}(b). $K_{\rm uniax}$ continuously decreases with rising temperature due to enhanced thermal fluctuations that eventually destroy magnetic order at $T_{\rm C}$. Nevertheless, its value $K_{\rm uniax} \approx 3.5\cdot 10^6$\,erg/cm$^3$ at the minimum temperature $T = 3\,{\rm K} \ll T_{C}$ can be considered as representing the ferromagnetic ground state of \MMS. 

The anisotropy field $H_{\rm a} = 2$\,T at $T = 3$\,K  was accurately reproduced with the SpinW numerical modeling [Fig.~\ref{fig:HF_ESR1}(b)] choosing in Hamiltonian~(\ref{eq:hamilton}) the single-ion anisotropy parameter $A^{\rm zz} = 0.1$\,meV. It is interesting to compare it with  $K_{\rm uniax}$ at the same temperature obtained from the analytical solutions based on Hamiltonian~(\ref{eq:free_energy}). The value of $A^{\rm zz}$ corrected for the demagnetization effect amounts to  $A^{\rm zz}_{\rm corr} = 0.07$\,meV ($1.12\cdot 10^{-16}$\,erg) and corresponds to the anisotropy energy of one Mn atom. Since there are six Mn atoms in the unit cell of \MMS, the respective magnetic anisotropy energy density should amount to  $6A^{\rm zz}_{\rm corr}/V_{\rm uc} = 2.8\cdot 10^6$\,erg/cm$^3$ with the unit cell volume $V_{\rm uc} = 238\,\AA^3$. This value is smaller than $K_{\rm uniax}$ and the discrepancy can be presumably attributed to the differences between the more simple phenomenological Hamiltonian~(\ref{eq:free_energy}) which treats a ferromagnet on the level of macroscopic magnetization and macroscopic anisotropy of the uniform solid whereas the more complex microscopic Hamiltonian~(\ref{eq:hamilton}) handles the system on the level of interatomic interactions.     

A significant magnetocrystalline anisotropy indicates an appreciable spin-orbit coupling in \MMS. Usually, the enhancement of SOC is attributed to the presence of heavy elements in the crystal structure covalently bonded to $3d$ transition metal ions, which is Sn in the present case. For instance, in the hexagonal kagome compound Mn$_3$Sn, density functional theory (DFT) calculations demonstrate an interesting interplay of spin and orbital contributions to the total magnetic moment induced by SOC on the Mn and Sn sites \cite{Nyari2019}. Remarkably, the easy-plane kagome ferromagnet Fe$_3$Sn$_2$, with the value of $K_{\rm uniax}$ very similar to which we have found for \MMS \cite{Prodan2025}, demonstrates a large anomalous Hall effect \cite{Kida2011,Wang2024} highlighting the significance of SOC for the nontrivial electronic properties of this kagome ferromagnet \cite{Zhang2024}. Moreover, very recent results of DFT and dynamical mean-field theory calculations have shown that hybridization between $d$-orbitals of Mn and $p$-orbitals of Sn, as well as SOC, indeed play a crucial role in the electronic properties of \MMS \cite{Sau2024}. The calculated electronic band structure features the Dirac point and a nodal line, which opens a gap in the presence of SOC that also generates finite Berry curvature along this line and gives rise to the anomalous Hall conductivity \cite{Ma2025}. It was shown that $p$-orbitals of Sn are involved in the Mn--Mn magnetic couplings, yielding competing ferro- and antiferromagnetic exchange interactions. 
Thus, in view of the results of the present FMR spectroscopic study, it is appealing to elucidate the microscopic origin of the magnetocrystalline anisotropy in \MMS by further {\it ab initio} relativistic electronic band structure calculations.  

\section{Conclusion}\label{sec:Conclusion}
In summary, we have studied ferromagnetic resonance response of the itinerant kagome magnet \MMS in the broad frequency, magnetic field, and temperature ranges to obtain insights into the origin of magnetic anisotropy of this compound, which yields the easy-plane type of FM order of Mn magnetic moments. From the analysis of the frequency and temperature dependences of the FMR modes using the numerical and analytical approaches of the linear spin wave theory, we could disentangle and quantify different contributions to the total anisotropy of \MMS. We have found that the extrinsic shape anisotropy due to the plate-like form of the crystals makes a minor contribution, whereas the intrinsic uniaxial magnetocrystalline anisotropy of the easy-plane type arising as a result of the spin-orbit coupling is significantly large and its energy density amounts to $K_{\rm uniax} \approx 3.5\cdot 10^6$\,erg/cm$^3$ at low temperature. Remarkably, it remains finite, although reduced to about $1\cdot 10^6$\,erg/cm$^3$ even at the critical temperature $T_{\rm C} \approx 300$\,K, suggesting that \MMS maintains at least short-range FM order upon a gradual transition to the paramagnetic state. Recent computational results suggest that SOC is essential for the nontrivial topological electronic band structure of \MMS, giving rise to such a remarkable phenomenon as the anomalous Hall effect \cite{Sau2024,Ma2025}. Our findings corroborate the significance of SOC for the properties of \MMS and call for first-principles band structure calculations of magnetocrystalline anisotropy in this compound to be then compared with experimental results. 

\section*{Appendix: Details on the samples' characterization}
\label{sec:appendix}

Compositions of the grown crystals of \MMS were estimated using energy-dispersive X-ray spectroscopy (EDXS, Quanta 250 FEG) on polished surfaces, yielding a ratio close to Mg:Mn:Sn = 1:6:6. To confirm the single-crystalline nature, X-ray diffraction (XRD) measurements were performed on the top surface of a crystal using a Rigaku SmartLab diffractometer with 9 kW Cu K$_\alpha$ radiation ($\lambda = 1.5418$ Å). The lattice parameters are estimated to be $a = b = 5.5236\,\text{\AA}$, $c = 9.0373\,\text{\AA}$, $\alpha = \beta = 90^{\circ}$, and $\gamma = 120^{\circ}$, with the volume of the unit cell 238.7887\,$\text{\AA}^3$.

\begin{figure*}
	\centering
	\includegraphics[width=1\textwidth]{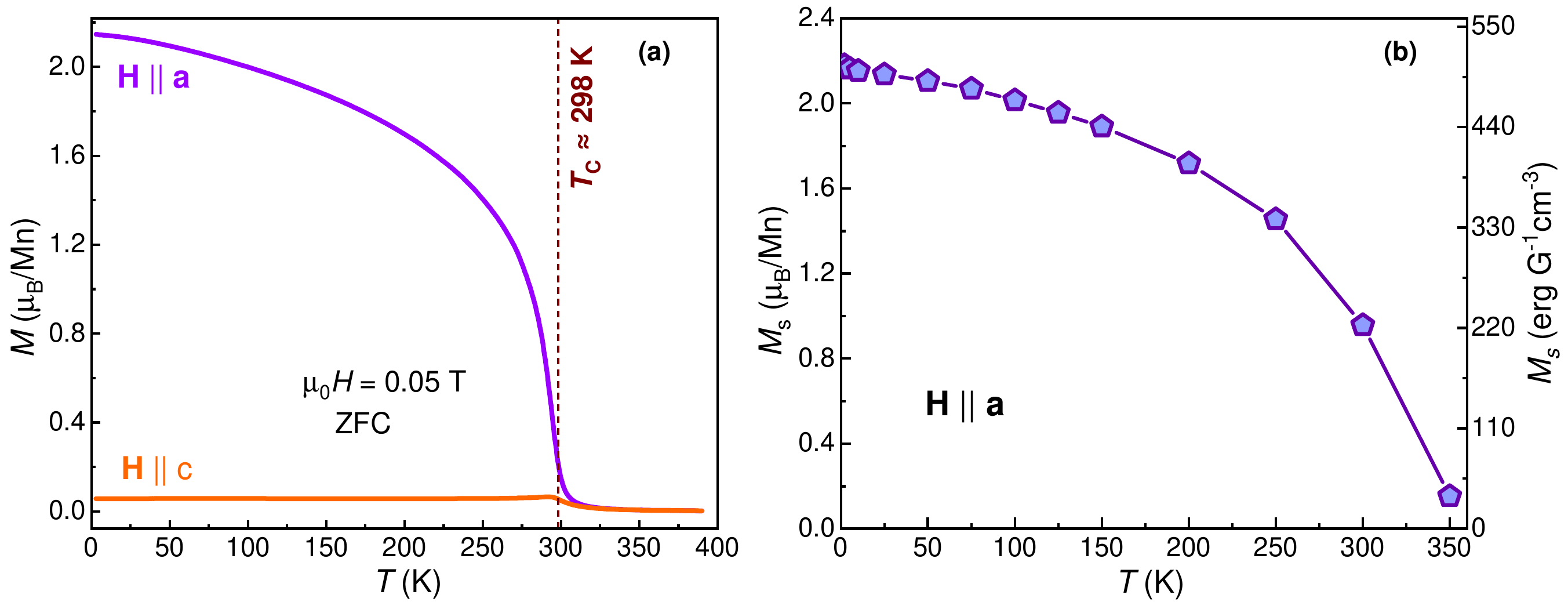}
	\renewcommand{\figurename}{Fig.}
	\caption{(a) Temperature-dependent magnetization measured under the zero-field-cooled (ZFC) protocol with an applied field of 0.05\,T for in-plane ($\mathbf{H}\parallel \mathbf{a}$) and out-of-plane ($\mathbf{H}\parallel \mathbf{c}$) directions. (b)~Temperature dependence of the saturation magnetization $M_{\rm s}(T)$ for $\mathbf{H}\parallel \mathbf{a}$.}
	\label{fig:magnetization}
\end{figure*}

Static magnetic measurements were carried out on polished crystals using the Vibrating Sample Magnetometer (VSM) option of a Quantum Design Physical Property Measurement System (DynaCool, PPMS-9T). Temperature-dependent magnetization $M(T)$ of \MMS was measured under an applied magnetic field of 0.05\,T along the \textit{a-} and \textit{c-}axes with zero-field-cooled (ZFC) and field-cooled warming (FCW) protocols which yielded practically identical results within the experimental error bars with only a slight low-temperature bifurcation for $\mathbf{H}\parallel \mathbf{c}$ indicating weak thermal irreversibility. The ZFC $M(T)$ curves are shown in Fig.~\ref{fig:magnetization}(a). A sharp upturn of $M(T)$ near 298\,K indicates the paramagnetic to ferromagnetic transition with \(T_{\mathrm{C}} \approx 298\,\mathrm{K}\). Below \(T_{\mathrm{C}}\), the in-plane response ($\mathbf{H}\parallel \mathbf{a}$) is significantly larger than the out-of-plane one ($\mathbf{H}\parallel \mathbf{c}$), consistent with easy-plane (\textit{ab}-plane) anisotropy. Fig.~\ref{fig:magnetization}(b) shows the temperature dependence of the saturation magnetization $M_{\rm s}(T)$ for $\mathbf{H}\parallel \mathbf{a}$ obtained from the $M(H)$ curves measured at selected temperatures. As can be seen there, the saturation moment decreases smoothly with increasing temperature, following the behavior expected for bulk ferromagnetic order.

\section*{Acknowledgments}
The authors would like to thank Joyal John Abraham for his technical assistance and helpful discussions. The authors gratefully acknowledge financial support by the Deutsche Forschungsgemeinschaft (DFG) through Project No. 499461434, and within the Collaborative Research Center SFB 1143 ``Correlated Magnetism - From Frustration to Topology'' (project-id 247310070), and the Dresden-W\"urzburg Cluster of Excellence (EXC 2147) ``ctd.qmat - Complexity and Topology in Quantum Matter'' (project-id 390858490). N.K. acknowledges financial support from the Science and Engineering Research Board (SERB), India, under Grant No. CRG/2021/002747, as well as funding from the Max Planck Society through the Max Planck–India Partner Group program. 

\section*{Declarations}

\textbf{Funding}: Deutsche Forschungsgemeinschaft, projects 499461434, 247310070, and 390858490; Science and Engineering Research Board (SERB), India, Grant No. CRG/2021/002747; Max Planck Society through the Max Planck–India Partner Group program.\\

\textbf{Conflict of interest}: The authors declare no conflict of interest.\\

\textbf{Ethics approval}: Not applicable.\\

\textbf{Availability of data and materials}: Data are available from the authors, based on reasonable requests.\\

\textbf{Authors' contributions}: R.P. performed FMR measurements, collected data, and performed the detailed analysis. K.D. and N.K. synthesized, characterized, and performed magnetization measurements of the crystals. All authors discussed the results. V.K., A.A., and B.B. developed the conception of the paper. V.K. wrote the first draft of the manuscript. All authors reviewed and edited the final version.  \\

\bibliography{Literature_MMS.bib}

\end{document}